# Direct mapping of nuclear shell effects in the heaviest elements

E. Minaya Ramirez[1,2], D. Ackermann[2], K. Blaum[3,4], M. Block[2], C. Droese[5], Ch. E. Düllmann[6,2,1], M. Dworschak[2], M. Eibach[4,6], S. Eliseev[3], E. Haettner[2,7], F. Herfurth[2], F.P. Heßberger[2,1], S. Hofmann[2], J. Ketelaer[3], G. Marx[5], M. Mazzocco[8], D. Nesterenko[9], Yu.N. Novikov[9], W.R. Plaß[2,7], D. Rodríguez[10], C. Scheidenberger[2,7], L. Schweikhard[5], P.G. Thirolf[11] and C. Weber[11]

[1]Helmholtz-Institut Mainz, 55099 Mainz, Germany

[2]GSI Helmholtzzentrum für Schwerionenforschung GmbH, 64291 Darmstadt, Germany

[3]Max-Planck-Institut für Kernphysik, 69117 Heidelberg, Germany

[4]Ruprecht-Karls-Universität, 69120 Heidelberg, Germany

[5]Ernst-Moritz-Arndt-Universität, 17487 Greifswald, Germany

[6]Johannes Gutenberg-Universität, 55099 Mainz, Germany

[7]Justus-Liebig-Universität, 35392 Gießen, Germany

[8]Dipartimento di Fisica and INFN Sezione di Padova, 35131 Padova, Italy

[9]Petersburg Nuclear Physics Institute, Gatchina, 188300 St. Petersburg, Russia

[10]Universidad de Granada, 18071 Granada, Spain

[11]Ludwig-Maximilians-Universität München, 85748 Garching, Germany

**Quantum-mechanical shell effects are expected to strongly enhance nuclear binding on an "island of stability" of superheavy elements. The predicted center at proton number $Z = 114$, 120, or 126 and neutron number $N = 184$ has been substantiated by the recent synthesis of new elements up to $Z = 118$. However the location of the center and the extension of the island of stability remain vague. High-precision mass spectrometry allows the direct measurement of nuclear binding energies and thus the determination of the strength of shell effects. Here, we present such measurements for nobelium and lawrencium isotopes, which also pin down the deformed shell gap at $N = 152$.**

Quantum-mechanical shell effects play a crucial role in determining the structure and the properties of matter. The electronic shell structure defines the architecture of the periodic table. An analogous effect leads to the so-called "magic nuclei," closed nucleon shells that lead to an enhanced binding of the atomic nucleus, that opposes Coulomb repulsion of protons and governs the landscape of the nuclear chart. The heaviest stable doubly magic nucleus is $^{208}$Pb with $Z = 82$, $N = 126$. The quest for the end of the periodic table and the northeast limit of the nuclear chart (Fig. 1) drives the search for even heavier magic nuclei.

In these superheavy elements (SHEs), nuclear shell effects are decisive for their mere existence. Without them, their nuclei would instantaneously disintegrate by spontaneous fission through Coulomb repulsion. A manifestation of these nuclear shell effects is an increase of the half-life by 15 orders of magnitude compared to liquid-drop-model predictions for nuclei around $N = 152$ [1]. Thus, SHEs are a prime testing ground for the understanding of shell effects and the character of the nuclear force.

Already in the late 1960s, about two decades after the introduction of the nuclear shell model [2,3], an "island of stability" of SHEs far from the known nuclei was predicted. Recent experimental evidence for the existence of isotopes of elements up to $Z = 118$ [4] has confirmed this concept, but the exact location and extension of this island are still unknown [5-7]. The presently known or claimed nuclides in the northeast end of the nuclear chart are shown in Fig.1. The blue shaded background indicates the gain in binding energy from shell effects. Regions of enhanced binding are predicted for the deformed magic nuclei at $N = 152$ and 162 around fermium ($Z = 100$) [1] and hassium ($Z = 108$) [8,9] and for spherical nuclei at $Z = 114$, $N = 184$.

Direct measurement of the strength of shell effects for SHE nuclei has been beyond experimental capabilities until now. It could only be derived either indirectly from a comparison of, e.g., experimental cross sections and half lives with predicted values, or from measured $Q_\alpha$ values, i.e. energy differences, in alpha decays. Here, we report the direct measurement of the neutron shell-gap by precision mass measurements on nobelium ($Z = 102$) and lawrencium ($Z = 103$) isotopes around $N = 152$. The results supply valuable information on the nuclear structure of SHEs, highly relevant for an improved prediction of the island of stability.

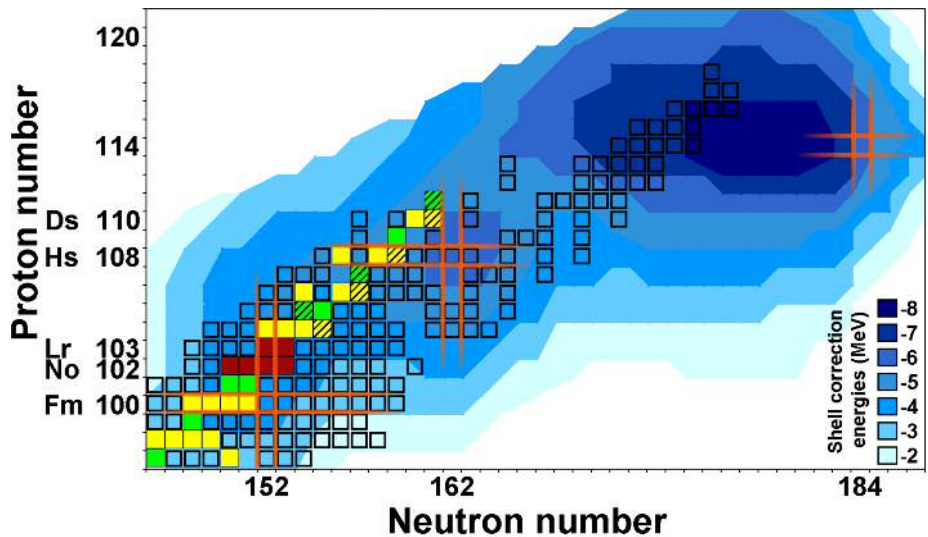

**Fig. 1** Chart of nuclides above berkelium ($Z = 97$). The blue background shows the calculated shell-correction energies [6]. The orange-shaded lines indicate known and predicted shell closures. The squares represent presently known or claimed nuclides. The nobelium and lawrencium isotopes whose masses are reported here are indicated by red squares. The yellow and green squares represent nuclides whose masses are determined by use of these new mass values, respectively, as anchor points in combination with experimental α-decay energies. Dashed squares show nuclides with unknown or ambiguous excited states. For details see text.

Mass spectrometry is a direct probe of nuclear stability, as the mass includes the total binding energy. Until recently, masses in the region of the heaviest elements could only be inferred via α-decay energies. For nuclides with even numbers of protons and neutrons, where the decay connects ground states, this approach is straightforward as the mass of the mother/daughter nucleus can be derived from the measured decay energy $E = \Delta mc^2$ and the mass of the daughter/mother nucleus, respectively. Although the uncertainties add up along decay chains, the masses of several nuclides between uranium ($Z = 92$) and copernicium ($Z = 112$) have been deduced this way [10].

However, in general, the situation is more complex as α decays preferably connect levels with identical configurations, whereas the ground-state configurations of mother and daughter nuclei usually differ for odd-$Z$ and/or odd-$N$ nuclides. These nuclei decay to excited states that in turn generally de-excite to the ground state by emission of photons or conversion electrons. Thus, the total decay energy is shared among the α particle, γ rays, and/or conversion electrons, i.e. the mere knowledge of the α-particle energy is insufficient. Unfortunately, for such nuclides unambiguous decay schemes, which would provide the needed information to obtain the true $Q_\alpha$ values, are rarely available. For many nuclides above fermium ($Z = 100$) the mass values are only extrapolated with uncertainties of several 100 keV [10].

In contrast, direct mass measurements provide absolute mass values and model-independent binding energies $E_B$ with no need for any ancillary information. Thus, by pinning down members of α-decay chains they can establish anchor points for a larger region in the chart of nuclei. In addition, accurate mass values define differential quantities such as the two-neutron separation energy $S_{2n}(N,Z) = E_B(N,Z) - E_B(N-2,Z)$ that are sensitive to quantum-mechanical shell effects and can uncover underlying nuclear-structure phenomena.

Direct high-precision mass measurements of short-lived nuclides are best performed with Penning traps [11,12] by determination of the cyclotron frequency $v_c = qB/(2\pi m)$ of the ion of interest with charge $q$ and unknown mass $m$ stored in a homogeneous magnetic field $B$. We have extended earlier $^{252-254}$No ($Z = 102$) measurements [13,14] to nuclei with an even larger number of protons and neutrons, $^{255,256}$Lr (lawrencium, $Z = 103$) and $^{255}$No/$^{256}$Lr ($N = 153$), respectively. The latter are found on the neutron-rich side of the $N = 152$ shell gap (see Fig. 1). Combining the new results with our previous measurements, we can now directly determine the strength of the $N = 152$ shell gap at $Z = 102$, the doorway to the SHEs.

SHIPTRAP [15] at GSI Darmstadt is a Penning-trap mass spectrometer installed behind the velocity filter SHIP, which is known in particular for the discovery of the elements with $Z = 107 - 112$ [16]. Nobelium and lawrencium isotopes were produced in fusion reactions of $^{48}$Ca projectile ions (accelerated to 218.4 MeV) with $^{206,207,208}$Pb and $^{209}$Bi targets. The product nuclei were separated from the primary beam by SHIP. Their kinetic energy was reduced from about 40 MeV by Mylar degrader foils and a 2-mg/cm$^2$ titanium entrance window to the SHIPTRAP gas cell. There, the particles were thermalized in 50-mbar ultrahigh-purity helium. They emerged mainly as doubly-charged ions, were cooled, bunched, and accumulated by a radiofrequency-quadrupole ion trap, and then transferred to a 7-Tesla double-Penning-trap system. In the first trap, the ions of interest were selected with a mass resolving power of up to $10^5$. In the second trap, the cyclotron frequency $v_c$ was determined with the time-of-flight ion-cyclotron-resonance method [17] (cf. Fig. 2). At least a few tens of ions must be detected to obtain a resonance. Thus, even for modern Penning traps, the extremely low production rates of exotic nuclei pose a major challenge for high-precision mass measurements. In the case of $^{256}$Lr, the present measurements required a long measurement time. With a cross section as low as $\sigma = 60$ nb [18], 93 hours were required for a resonance based on 48 detected ions (see Fig. 2).

The mass of the ion of interest was determined by comparing its cyclotron frequency ($v_c$) with that of the reference ion $^{133}$Cs$^+$ ($v_{c,ref}$), with well-known mass [10] and a mass-to-charge ratio close to the doubly-charged nobelium and lawrencium ions. The statistical uncertainty depends on the number of detected ions per resonance and the Fourier-limited resolution $\Delta v_c \approx 1/t$, which is inversely proportional to the excitation time t that a radiofrequency signal is applied to excite the cyclotron motion of the stored ions. In addition, a systematic uncertainty of $4.5 \cdot 10^{-8}$ (based on the residual scattering observed in cross-check measurements with carbon-cluster ions [19]) has been taken into account.

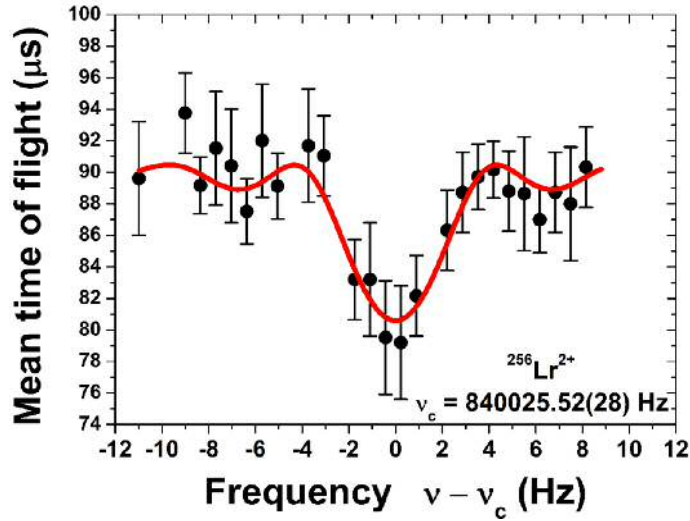

**Fig. 2.** Time-of-flight ion-cyclotron-resonance of doubly charged $^{256}$Lr$^{2+}$ (t = 200 ms). The curve is a fit of the expected line shape to the data. Error bars, ±1 s.d.

The resulting frequency ratios are given in Table 1. In addition to the confirmation for the even-Z nobelium isotopes $^{252}$No and $^{254}$No [14] and a further reduction of their uncertainties, accurate mass values were also determined for the odd-Z element lawrencium. Overall, the decay-energy based values [10] agreed, within their fairly large systematic uncertainties, with our new direct mass measurements of $^{255}$No, $^{255}$Lr, and $^{256}$Lr.

**Tab. 1:** Measured frequency ratios and resulting mass-excess ($M_{exc}$ = $M$[atomic mass] – $A$[atomic mass number]*$u$ [atomic mass unit]), values for the nuclides investigated in this work as well as theoretical predictions from several models [6,23,24,25]. The Z and N values of the predicted center of the island of stability by each model are indicated.

| Isotope | Frequency ratio | Mass excess (keV) | | | | |
|---|---|---|---|---|---|---|
| | $\nu_{c,ref}/\nu_c$ | Experiment SHIPTRAP | FRDM [2] Z = 114 N = 184 | MPS-03 [24] Z = 114 N = 184 | TW-99 [25] Z = 120 N = 172 | SkM* [23] Z = 126 N = 184 |
| $^{252}$No | 0.94837684(7) | 82870(16) | 82200 | 82950 | 91566 | 95308 |
| $^{253}$No | 0.95214494(5) | 84356(13) | 83840 | 84190 | - | 96396 |
| $^{254}$No | 0.95590852(6) | 84726(14) | 84050 | 84700 | 93646 | 96989 |
| $^{255}$No | 0.95967902(6) | 86808(15) | 86550 | 86700 | - | 98615 |
| $^{255}$Lr | 0.95969174(6) | 89958(16) | 89300 | 89920 | - | 102326 |
| $^{256}$Lr | 0.9634610(3) | 91746(83) | 91420 | 91690 | - | 103727 |

The SHIPTRAP masses provide anchor points of α-decay chains and thus affect the masses of many other nuclides. The mass of $^{255}$No determines the masses of its daughters linked by α-decay ($^{247}$Cf) and electron capture ($^{247}$Bk). Until recently, the nuclear spectroscopy data of the decay chain from $^{270}$Ds (darmstadtium, Z = 110) to $^{254}$No had been incomplete [20]. However, the recent discovery of an α-decay branch of $^{262}$Sg [21] provides now the missing link. Thus, the mass excess of $^{270}$Ds was experimentally established with an uncertainty of 40 keV. It is the highest-Z anchor point near the deformed doubly-magic nucleus $^{270}$Hs (Z = 108, N = 162) [8] (cf. Fig. 1). Thus, our measurements impact nuclides in a region reaching into the realm of the SHEs.

These high-precision mass values can be used to benchmark nuclear models [22] and can be extended to the very heavy nuclides. The centre of the island of stability around $N$ = 184 is experimentally still out of reach, and reliable theoretical predictions are crucial for directing future experimental efforts. The masses measured at SHIPTRAP provide direct, model-independent values of the total binding energies and precise values for derived quantities such as the shell gap parameter [23],

$$\delta_{2n}(N,Z) = S_{2n}(N,Z) - S_{2n}(N+2,Z) = -2 M_{exc}(N,Z) + M_{exc}(N-2,Z) + M_{exc}(N+2,Z),$$

that theoretical predictions can be confronted with or used to calculate other quantities. $\delta_{2n}(N,Z)$ is a sensitive indicator of shell closures and a measure of their strength, and appropriate for examining shell stabilization for $N$ = 152. Because the uncertainties of our results are well below the differences among the predictions by the different models they allow us to test their predictive power. We have selected as representative examples two microscopic-macroscopic models, namely the global FRDM [6] and the approach by Muntian et al. [24] that is optimized locally for SHEs and served for producing the blue-shaded contour plot underlying Fig. 1, as well as a self-consistent mean-field model using the Skyrme-Hartree-Fock effective interaction SkM* [23] and, furthermore, a relativistic mean-field model using the effective interaction TW-99 (only even-even nuclei) [25].

Microscopic-macroscopic approaches are based on the liquid-droplet model with modifications by a microscopic shell-correction energy [26] crucial for the existence of SHEs. These models describe various nuclear properties across the nuclear chart and predict the island of stability at $Z = 114$ and $N = 184$. Self-consistent mean-field models use energy-density functionals based on effective nucleon-nucleon interactions. Different effective interactions (e.g. Skyrme and Gogny forces) have been developed, which lead to different predictions for the location of the island of stability, e.g. around $Z = 126$, $N = 184$ for the effective interaction SkM*. In recent years relativistic mean-field models have been applied. In the TW-99 parameterization the spherical SHEs shells are predicted at $Z = 120$, $N = 172$.

Fig. 3 shows a comparison between the predicted and experimental $\delta_{2n}(N,Z)$ values around $N = 152$ for nobelium and lawrencium. Compared to the drastic deviation of the predicted mass excess of up to about 10 MeV from the experimental data (cf. Tab. 1), the agreement of $\delta_{2n}(N,Z)$ is significantly better, with deviations below about 500 keV. Not surprisingly, the local model by Muntian et al. agrees best with the data. All models show a similar trend with increasing neutron number and a peak indicating the deformed $N = 152$ shell except for the TW-99 parameterization. However, the strength of the shell gap differs strongly. From our data, we obtain $\delta_{2n}(152,102) = 1242(16)$ keV. The macroscopic-microscopic models predict a larger neutron shell gap of 1310 keV (Muntian et al.) and 1520 keV (FRDM), whereas the self-consistent mean-field model, SkM*, yields a much smaller and more smeared out shell gap (982 keV). For the TW-99 parameterization, where only data for even-even nuclei are available, a clear shell gap for $Z = 102$ at $N = 154$ of 1375 keV is obtained, also larger than observed experimentally. The difference in values for the size of the shell gap or even a shift from $N = 152$ to $N = 154$ for one of the selected models illustrates the sensitivity of the different predictions on the used parameterization.

The values of $\delta_{2n}(N,Z)$ obtained from our data confirm the existence of a region of enhanced shell stabilization for nuclides above the doubly-magic $^{208}$Pb. In particular for $Z = 102$ our data pin down the shell gap at $N = 152$ accurately. The findings are corroborated for $Z = 103$. Thus, our mass measurements have established, with high accuracy, a shell gap at $N = 152$ - a major input for theoretical models which have not yet converged to a common prediction of the location and the strength of the shell closures in SHEs.

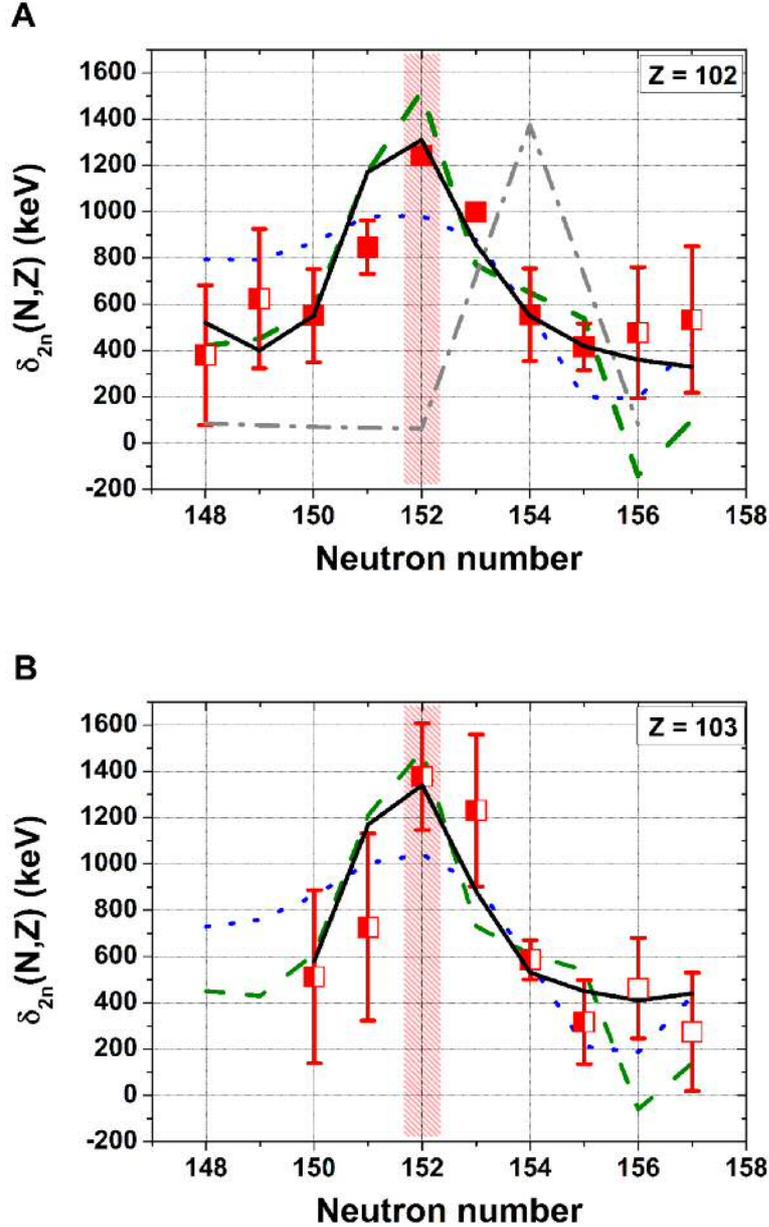

**Fig. 3.** Experimental shell gap $\delta_{2n}(N,Z)$ (red squares and error bars) and theoretical predictions (dashed green line: Möller [6]; dotted blue line: SkM* [23]; black line: Muntian et al. [24]; dash-dotted gray line: TW-99 [25], only for A) for nobelium (A) and lawrencium (B). For further details see text. Experimental values have been calculated using one (semi-filled red squares) or two (filled red squares) masses from this work. Values from the AME 2003 are shown by open red squares. The shaded area at $N = 152$ indicates the position of the deformed neutron shell gap. Error bars, ±1 s.d.

For the first time the masses of $^{255}$No and $^{255,256}$Lr have been measured directly. Mass uncertainties as low as 15 keV have been achieved and $^{256}$Lr is the heaviest nuclide and the one with the smallest production rate ever investigated at an on-line Penning-trap mass spectrometer. The combination of the present results with spectroscopic data fixes the masses of nuclides as heavy as $^{270}$Ds ($Z$ = 110). Moreover, the accurate experimental binding energies allow mapping the shell effect at $N$ = 152. Thus, our mass measurements on isotopes with production cross sections of the order of only tens of nanobarn pave the way to a better understanding of the superheavy elements.

**References and Notes:**

1. Z. Patyk and A. Sobiczewski, Nucl. Phys. A 491 (1989) 267
2. M. Göppert-Mayer, Phys. Rev. 74 (1948) 235
3. M. Göppert-Mayer and J. H. D. Jensen, *Elementary Theory of Nuclear Shell Structure (Wiley, New York)*, 1955
4. Yu.Ts.Oganessian et al., Radiochim. Acta 99 (2011) 429
5. S. Cwiok et al., Nature 433 (2005) 705
6. P. Möller and J.R. Nix, At. Data Nucl. Data Tab. 59 (1995) 185
7. M. Bender et al., Phys. Rev. C 60 (1999) 034304
8. J. Dvorak et al., Phys. Rev. Lett. 97 (2006) 242501
9. A. Sobiczewski et al., Prog. Part. Nucl. Phys. 58 (2007) 292
10. G. Audi et al., Nucl. Phys. A 729 (2003) 337
11. L. Schweikhard and G. Bollen, (eds.) special issue of Int. J. Mass Spectrom. 251 (2006)
12. K. Blaum, Phys. Rep. 425 (2006) 1
13. M. Block et al., Nature 463 (2010) 785
14. M. Dworschak et al., Phys. Rev. C 81 (2010) 064312
15. M. Block et al., Eur. Phys. J. D 45 (2007) 39
16. S. Hofmann and G. Münzenberg, Rev. Mod. Phys. 72 (2000) 733
17. G. Gräff et al., Z. Phys. 222 (1969) 201
18. F.P. Heßberger et al., Eur. Phys. J. D 45 (2007) 33
19. A. Chaudhuri et al., Eur. Phys. J. D 45 (2007) 47
20. S. Hofmann et al., Eur. Phys. J. A 10 (2001) 5
21. D. Ackermann et al., GSI Ann. Rep. 2010 - GSI Rep. 2011-1 (2011) 200
22. D. Lunney et al., Rev. Mod. Phys. 75 (2003) 1021
23. K. Rutz et al., Phys. Rev. C 56 (1997) 238
24. I. Muntian et al., Act. Phys. Pol. B 34 (2003) 2073

25. W. Zhang et al., Nucl. Phys. A 753 (2005) 106

26. V.M. Strutinsky, Nucl. Phys. A 95 (1967) 420

**Acknowledgments:** The project was supported in part by the Helmholtz-Institut Mainz (HIM), the GSI Helmholtzzentrum für Schwerionenforschung GmbH, the German Federal Ministry of Education and Research (BMBF), the Max-Planck Society, the Russian Minobrnauki, and the ExtreMe Matter Institute (EMMI). Yu.N. acknowledges support by EMMI.